\documentstyle[emulateapj-djb,epsfig]{article}   
\slugcomment{}
\begin{document}

\title{350 Micron Dust Emission from High Redshift Objects}

\author{Dominic J. Benford\altaffilmark{1}, Pierre
  Cox\altaffilmark{2}, Alain Omont\altaffilmark{3}, Thomas G.
  Phillips\altaffilmark{1}, Richard G.  McMahon\altaffilmark{4}}

\altaffiltext{1}{California Institute of Technology, MS 320-47,
  Pasadena, CA 91125}
\altaffiltext{2}{Institut d'Astrophysique Spatiale, B\^at. 121,
  Universit\'e de Paris XI, F-91405 Orsay, France}
\altaffiltext{3}{Institut d'Astrophysique de Paris, CNRS, 98bis
  boulevard Arago, F-75014 Paris, France}
\altaffiltext{4}{Institute of Astronomy, Madingley Road, Cambridge CB3
  0HA, UK}

\begin{abstract}
  
  We report observations of a sample of high redshift sources ($1.8
  \lesssim z \lesssim 4.7$), mainly radio-quiet quasars, at 350$\,\mu$m
  using the SHARC bolometer camera at the Caltech Submillimeter
  Observatory.  Nine sources were detected ($\rm \ga 4\sigma$) and
  upper limits were obtained for 11 with 350$\,\mu$m flux density limits
  (3$\,\sigma$) in the range 30$\,$-$\,125 \,$mJy.  Combining
  published results at other far-infrared and millimeter wavelengths
  with the present data, we are able to estimate the temperature of
  the dust, finding relatively low values, averaging 50$\,$K.  From the
  spectral energy distribution, we derive dust masses of a few $10^8
  \, h_{100}^{-2} M_{\odot}$ and luminosities of 4-33$\times 10^{12} \,
  h_{100}^{-2} L_{\odot}$ (uncorrected for any magnification) implying
  substantial star formation activity. Thus both the temperature and
  dust masses are not very different from those of local ultraluminous
  infrared galaxies. For this redshift range, the 350$\,\mu$m
  observations trace the 60-100$\,\mu$m rest frame emission and are thus
  directly comparable with IRAS studies of low redshift galaxies.

\end{abstract}
          

\section{Introduction}

The study of molecular gas and dust is a significant observational
tool for probing the physical conditions and star formation
activity in local galaxies. Recent advances in observational
techniques at submillimeter and millimeter wavelengths now permit
such studies to be made at cosmological distances.

As a result of the IRAS survey many local objects are recognized as
containing a large mass of dust and gas, such that the objects may be
more luminous in the far-infrared (FIR) than in the optical.  The
question as to how many such objects there may be at cosmological
distances, and whether they can account for the recently discovered
FIR cosmic background (\cite{Puget96}; \cite{Fixsen98};
\cite{Hauser98}), is attracting much interest and spawning new
instrument construction and new surveys (\cite{Hughes98};
\cite{Ivison98}; \cite{Kawara98}; \cite{Puget98}; \cite{Barger98};
\cite{Lilly99}).  Existing
submillimeter cameras on ground-based telescopes are not yet sensitive
and large enough to detect distant objects at 350$\,\mu$m in arbitrary
blank fields, e.g., the initial Caltech Submillimeter Observatory
(CSO) SHARC survey which achieved 100$\,$mJy ($1\,\sigma$) over about
10 square arcminutes (\cite{Phillips}).  However, such cameras can measure
the 350$\,\mu$m flux of objects of known position.  A step forward in
the field was the recognition of IRAS~F10214+4724 as a high redshift
object ($z=2.286$) by Rowan-Robinson et al. (1991).  However, it has
proved difficult to find many such objects to study.  On the other
hand, quasars have sometimes proved to exist in the environs of dusty
galaxies (\cite{Haas98}; \cite{Lewis98}; \cite{Downes98}).  Omont et
al. (1996b) have shown by means of a $1300\,\mu$m survey of radio-quiet
quasars that the dust emission at high redshifts can be detected in a
substantial fraction of their project sources.

In this letter we present measurements at 350$\,\mu$m towards a sample
of 20 sources with redshifts $1.8\lesssim z\lesssim 4.7$ which were
selected from different surveys and studies (references are given in
Tables \ref{detections} \& 2).  The sources had
previously been detected at longer wavelengths, and are predominantly
from the work of Omont et al. (1996b) and Hughes, Dunlop \& Rawlings
(1997).  A wavelength of 350$\,\mu$m roughly corresponds to the peak
flux density of highly redshifted ($z\simeq3$) dust emission of
objects with temperatures of 40 to 60$\,$K.  Together with
measurements at longer wavelengths, it strongly constrains the dust
temperature and, hence, the dust mass and, especially, the luminosity
of the object. Some of these results were first presented by Benford
et al. (1998).  In this paper, we assume ${\rm H}_0 = h_{100} \times
100 \, \rm km \, s^{-1} \, Mpc^{-1}$ and $\Omega_0 = 1$.


\begin{table*}[bhpt]
\caption{Sources detected at 350$\,\mu$m and derived properties.
\label{detections}} 
\begin{center}
 \footnotesize{
\begin{tabular}{llccccccc}
\hline 
\hline \\[-0.2cm] 
\multicolumn{1}{c}{Source} & \multicolumn{1}{c}{$z$} 
        & \multicolumn{1}{c}{R.A.} & \multicolumn{1}{c}{Dec.} 
        & \multicolumn{1}{c}{Flux Density} & \multicolumn{1}{c}{T$\rm _{dust}$}
        & \multicolumn{1}{c}{Dust Mass\tablenotemark{\ddagger}} & 
       \multicolumn{1}{c}{Luminosity\tablenotemark{\ddagger}} & Ref. \\
  & & \multicolumn{2}{c}{(B1950.0)} & (mJy, $\pm1\,\sigma$)& (K) & 
 (10$^{8} h_{100}^{-2}$ M$_{\odot}$) 
 & (10$^{12} h^{-2}_{100}$ L$_{\odot}$) & \\[0.2cm]
\hline \\[-0.2cm] 
BR1202$-$0725  & 4.69 & 12 02 49.3 & $-$07 25 50 &  106$\pm$7\phn & 50$\pm$7
        & 4.0$ ^{+0.9}_{-0.8}$ & 14.9$ ^{+0.8}_{-0.7}$ & 1,2,3,4 \\[0.1cm]
BRI1335$-$0417 & 4.41 & 13 35 27.6 & $-$04 17 21 &   52$\pm$8  & 43$\pm$6 
        & 3.6$ ^{+0.5}_{-0.8}$ & \phn6.0$ ^{+0.9}_{-0.5}$ & 1,2,5 \\[0.1cm]
HM0000$-$263    & 4.10 & 00 00 49.5 & $-$26 20 01 & 134$\pm$29  & 60\tablenotemark{\dagger} 
        & 2.0$ ^{+0.2}_{-0.2}$ & 20.7$ ^{+1.5}_{-1.5}$ & 2,6 \\[0.1cm]
4C41.17        & 3.80 & 06 47 20.8 & $+$41 34 04 &  37$\pm$9   & 52$\pm$6 
        & 1.0$ ^{+0.1}_{-0.1}$ & \phn4.3$ ^{+1.4}_{-1.3}$ & 7,8 \\[0.1cm]
PC2047$+$0123  & 3.80 & 20 47 50.7 & $+$01 23 56 & 80$\pm$20  &
      50\tablenotemark{\dagger}  & 2.3$ ^{+0.6}_{-0.6}$ & \phn8.5$^{+2.2}_{-1.6}$ & 9,10 \\[0.1cm]
Q1230+1627    & 2.74 & 12 30 39.4 & $+$16 27 26 & 104$\pm$21  &
     49$\pm$12  & 2.5$ ^{+0.4}_{-0.4}$  & 8.2$^{+2.0}_{-2.0}$ & 2,11  \\[0.1cm]
Q0100+1300   & 2.68 & 01 00 33.4 & $+$13 00 11 & 131$\pm$28  & 68$\pm$5 
        & 1.2$ ^{+0.5}_{-0.3}$ & 23.8$ ^{+0.2}_{-5.8}$ &  12,13\\[0.1cm]
H1413+117\tablenotemark{\ddagger}      & 2.54 & 14 13 20.1 & $+$11 43 38 & 293$\pm$14  & 45$\pm$7 
        & 8.9$ ^{+1.9}_{-1.9}$ & 17.7$ ^{+1.3}_{-0.6}$ & 14,15 \\[0.1cm]
F10214+4724\tablenotemark{\ddagger}    & 2.28 & 10 21 31.1 & $+$47 24 23 & 383$\pm$51 & 55$\pm$3
        & 5.5$ ^{+1.4}_{-1.4}$ & 32.8$  ^{+6.5}_{-5.3}$ & 16 \\[0.1cm]
\hline \\[-0.2cm]
\end{tabular} 
}
\end{center}\vspace{-0.5cm}
\tablenotetext{\dagger}{
Because of a lack of sufficient data at other wavelengths, the
temperatures of these sources are poorly constrained.  The available
upper limits constrain HM0000$-$263 to $50\lesssim \rm T_{dust}
\lesssim 75$K, so we adopt 60K; PC2047+0.123 is essentially
unconstrained, so we adopt 50K as that is roughly the median value for the dust temperature.}

\tablenotetext{\ddagger}{The dust masses and luminosities
listed above have not been corrected for any lensing amplification.
The probable intrinsic values may be found by
dividing by $A=7.6$ for H$\,$1413+117 (\cite{Alloin97}) and $A=13$ for
F10214+4724 (\cite{Downes95}).}

\tablerefs{
1. Storrie-Lombardi et al. (1996); [2.] Omont et al. (1996b);
[3.] McMahon et al. (1994); [4.] Isaak et al. (1994);
[5.] Guilloteau et al. (1997); 6. Schneider, Schmidt \& Gunn (1989);
7. Chambers, Miley \& van Bruegel (1990); [8.] Hughes, Dunlop \& Rawlings(1997);
9. Schneider, Schmidt \& Gunn (1994); [10.] Ivison (1995);  
11. Foltz et al. (1987); 12. Steidel \& Sargent (1991); 
[13.] Guilloteau et al. (1999); 14. Hazard et al. (1984); 
[15.] Barvainis, Coleman \& Antonucci (1992);
[16.] Rowan-Robinson et al. (1993) \\
Note: the references in brackets refer to the observations shown in Fig.~2.}  
\normalsize

\end{table*}

\section{Observations and Results}

The measurements were made during a series of observing runs in 1997
February and October and 1998 January and April with the 10.4$\,$m
Leighton telescope of the CSO on the summit of Mauna Kea, Hawaii,
during excellent weather conditions, with 225$\,$GHz atmospheric
opacities of $\lesssim0.05$ (corresponding to an opacity of
$\lesssim1.5$ at 350$\,\mu$m).  We used the CSO bolometer camera, SHARC,
described by Wang et al. (1996) and Hunter, Benford, \& Serabyn
(1996). It consists of a linear 24 element close-packed monolithic
silicon bolometer array operating at 300$\,$mK.  During the
observations, only 20 channels were operational. The pixel size is
$5''$ in the direction of the array and $10''$ in the cross direction.
The weak continuum sources were observed using the pointed observing
mode with the telescope secondary chopping in azimuth by 30$\rm
^{\prime\prime}$ at a rate of 4$\,$Hz.  The telescope was also nodded
between the on and the off beams at a rate of $\sim$0.1Hz.  The
point-source sensitivity of SHARC at 350$\,\mu$m is $\rm \sim 1 \,
Jy/\sqrt{Hz}$ and the beam size is $\sim9''$ FWHM. All measurements
were made at $350\mu$m with the exception of H$\,$1413+117 which was
also observed at $450\mu$m.

Pointing was checked regularly on nearby strong galactic sources which
also served as secondary calibrators, and was found to be stable with
a typical accuracy of $ \lesssim 3^{\prime\prime}$.  The planets Mars,
Saturn and Uranus served as primary flux calibrators. The absolute
calibration was found to be accurate to within 20\%.  Repeated
observations of H$\,1413+117$ and F$\,10214+4724$ confirmed a relative
flux accuracy of $\sim20\%$.  The data were reduced using the CSO
BADRS software package.  Typical sensitivities ($\rm 1 \sigma$) of
$\sim20$ mJy were achieved, after $\sim 2500\,$s of on--source
integration time.

Nine sources were detected at levels of $4\,\sigma$ and above, as
outlined in Table~\ref{detections}.  Included are the $z\,>\,4$
quasars BR$\,$1202$-$0725, BRI$\,$1335$-$0417 and HM$\,$0000$-$263.
Except for the Cloverleaf (H$\,$1413+117; \cite{Barvainis92}), the
present measurements are the first reported detections for high
redshift quasars at 350$\,\mu$m.  Many of the sources were measured
three or more times providing both consistency checks and improvements
in the accuracy of the flux densities. The two strongest sources,
H$\,$1413+117 and IRAS~F10214+4724, were often measured before
starting the long ($\sim$2-3 hours) integrations on the weaker
sources.

As an illustration of the data quality, Figure~1 shows the
350$\,\mu$m CSO--SHARC measurement towards BR$\,$1202$-$0725 at
$z\,=\,4.69$.  This measurement corresponds to a total of 4 hours
integration on source.  The source is centered at offset zero. The
other channels provide a measure of the neighboring blank sky emission
and a reference for the quality of the detection.

11 sources with redshifts between 1.8 and 4.5 were not detected at
350$\,\mu$m, with flux density upper limits at the 3$\,\sigma$ level of
30$\,$-$\,125\,$mJy.  Table~2 lists their names,
redshifts, 350$\,\mu$m flux density measurements with $\pm1\,\sigma$
errors, and a 3$\,\sigma$ upper limit to their luminosities (see below).

        \bigskip
\centerline{\epsfig{file=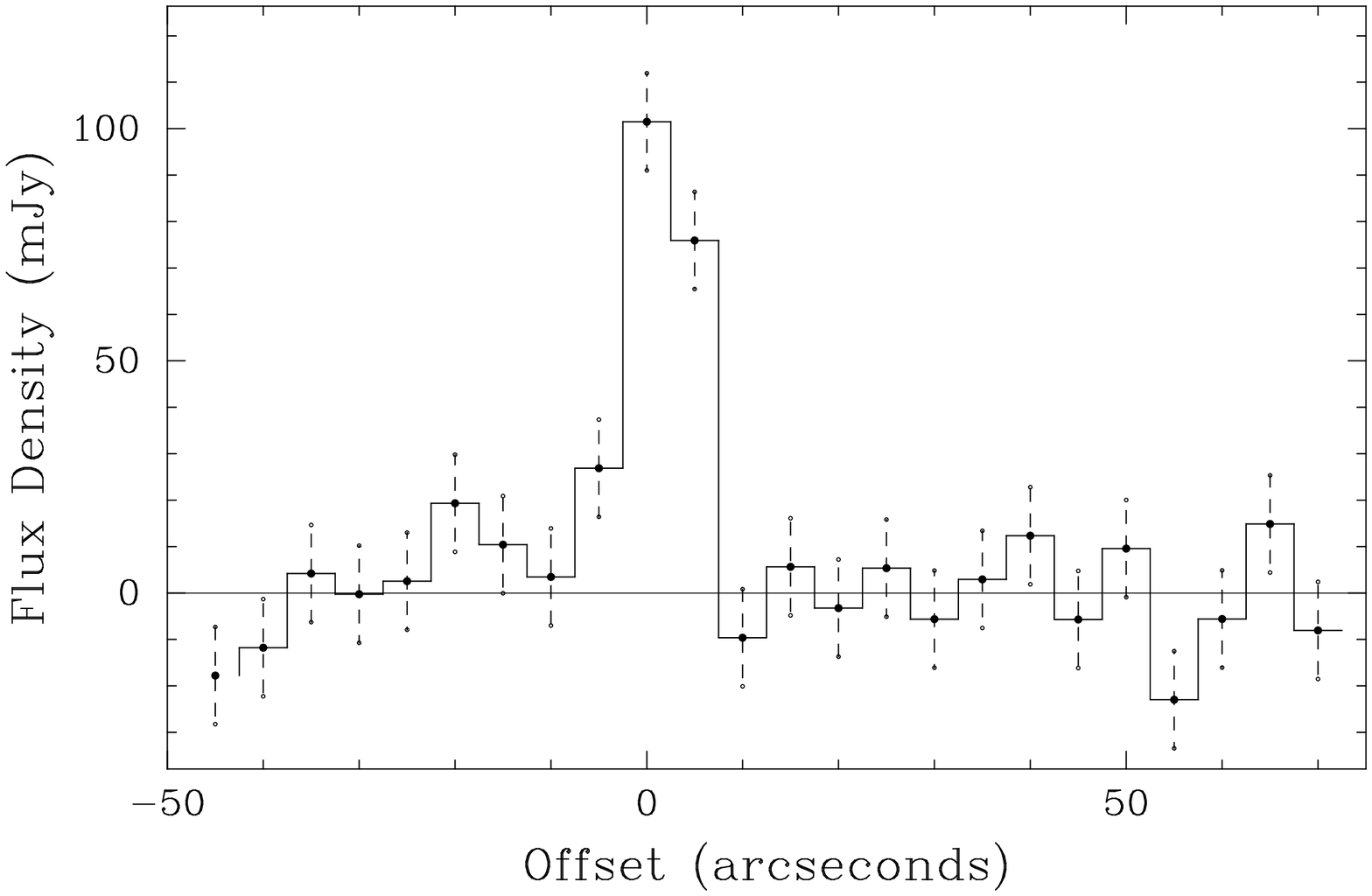, width=2.1in,angle=-90}}
        \medskip
        {\noindent\small Fig. 1 -- 
The 350$\,\mu$m flux density measured in the 
   bolometers of the SHARC array towards BR1202$-$0725 at $z$=4.69.
   Offsets are given in arcsec with respect to the reference channel
   number 10. The source is centered at offset zero. Emission is also 
   seen in the two neighboring pixels because the bolometers
   sample the diffraction pattern of the telescope with a Nyquist sampling.}
        \bigskip

        \bigskip
         \begin{center}
        \small
        {\sc TABLE 2} \\
        {\sc Sources with upper limits at 350$\,\mu$m}
        \vskip 1ex
\begin{tabular}{llccc}
\hline 
\hline \\[-0.2cm]
\multicolumn{1}{c}{Source} & \multicolumn{1}{c}{$z$} 
& \multicolumn{1}{c}{S$_{350}$}  & L$_{\rm FIR}$ ($3\,\sigma$) & Ref. \\
\multicolumn{2}{c}{ } & \multicolumn{1}{c}{(mJy, $\pm1\,\sigma$)} &
(10$^{12} h^{-2}_{100}$ L$_{\odot}$) & \\[0.2cm]
\hline \\[-0.2cm]
BR2237$-$0607   & 4.56 & \phs$\phn5\pm15$ & $<6$ & 1   \\
BRI0952$-$0115  & 4.43 & \phs$\phn8\pm22$ & $<8$ & 1,2 \\
PSS0248+1802    & 4.43 &     $-75\pm22$   & $<9$ & 3   \\
BR1117$-$1329   & 3.96 & \phs$27\pm13$    & $<4$ & 1   \\
Q0302$-$0019    & 3.28 & \phs$34\pm21$    & $<5$ & 4   \\
Q0636$+$680     & 3.18 & $-123\pm38\phn$  & $<9$ & 5   \\ 
Q2231$-$0015    & 3.01 & $-65\pm24$       & $<6$ & 4   \\
MG0414$+$0534   & 2.64 & $-24\pm35$       & $<8$ & 6   \\ 
Q0050$-$2523    & 2.16 & \phs$69\pm42$    & $<9$ & 4   \\
Q0842$+$3431    & 2.13 & \phs$16\pm10$    & $<2$ & 1   \\
Q0838$+$3555    & 1.78 & \phs$39\pm19$    & $<4$ & 1   \\[0.2cm]
\hline \\[-0.2cm]
\end{tabular}
         \end{center}
\vspace*{3ex}{\parbox{3.3in}{\hskip 1em \rm\footnotesize References. --- 
 1. Omont et al. (1996b); 2. Guilloteau et al. (1999); 
3. Kennefick et al. (1995); 4. Hewett, Foltz \& Chaffee (1995); 
5. Sargent, Steidel \& Boksenberg (1989); 6. Barvainis et al. (1998) }\par}
\smallskip 
        \normalsize

\section{Discussion}

Figure~\ref{seds} displays the spectral energy distributions of the 
six sources detected at 350$\,\mu$m for which fluxes at two or more 
other wavelengths are available from the literature. In the following, 
we will first comment on individual sources and then discuss the physical
properties of the objects.

The two radio-quiet, $z>4$ quasars, BR$\,$1202$-$0725 and
BRI$\,$1335$-$0417, are exceptional objects with large masses of gas
($\rm \sim 10^{11} \, M_{\odot}$) which have been detected in CO by
Omont et al. (1996a) and Guilloteau et al. (1997). They both are
clearly detected at 350$\,\mu$m with flux densities of $106\pm7$ and
$52\pm8\,$mJy, respectively.  However, BRI$\,$0952$-$0115, which is
the third $z>4$ quasar in which CO has been measured
(\cite{Guilloteau99}), is not detected at 350$\,\mu$m at a 3$\,\sigma$
level of 65$\,$mJy. This upper limit is consistent with the weak flux
density of BRI$\,$0952$-$0115 at 1.3~mm, $\rm 2.8 \pm 0.6 \, mJy$
(\cite{Omont96b}; \cite{Guilloteau99}), and a temperature of $\rm \sim
50 \, K$ (see below). The radio-quiet quasar HM$\,$0000$-$263 at
$z=4.11$, which was not detected at $\rm 1.25 \, mm$ using the 30m
telescope (\cite{Omont96b}), due to its low declination, shows a
large flux density at 350$\,\mu$m (134$\pm$29$\,$mJy).
Measurements at other wavelengths would be useful to further constrain
the properties of this object.

The detection of the $z=3.8$ radiogalaxy 4C41.17 with a flux density
of $37\pm9\,$mJy at 350$\,\mu$m is one of the most sensitive
measurements of this study. This sensitivity was reached after only
3/4 of an hour of on-source time and defines the limits which can be
achieved with SHARC in the pointed observing mode under excellent
weather conditions.  A marginal detection ($4\,\sigma$) was achieved
at 350$\,\mu$m of PC$\,$2047+0123, a $z=3.80$ quasar studied by
\cite{Ivison95}.  Finally, the 350$\,\mu$m flux density of the
Cloverleaf (H$\,$1413+117) is significantly higher than the value
published by Barvainis, Antonucci, \& Coleman (1992), i.e.
293$\pm14\,$mJy as compared to 189$\pm56\,$mJy. We have also obtained
for the Cloverleaf a 450$\,\mu$m flux density of $\rm 226\pm34 \, mJy$
in excellent agreement with the measurement of $\rm 224\pm38\,mJy$ at
$438\,\mu$m of Barvainis, Antonucci, \& Coleman (1992), as shown in
Figure~\ref{seds}.

        \setcounter{figure}{1}
\begin{figure*}[tbh]
\centerline{\epsfig{file=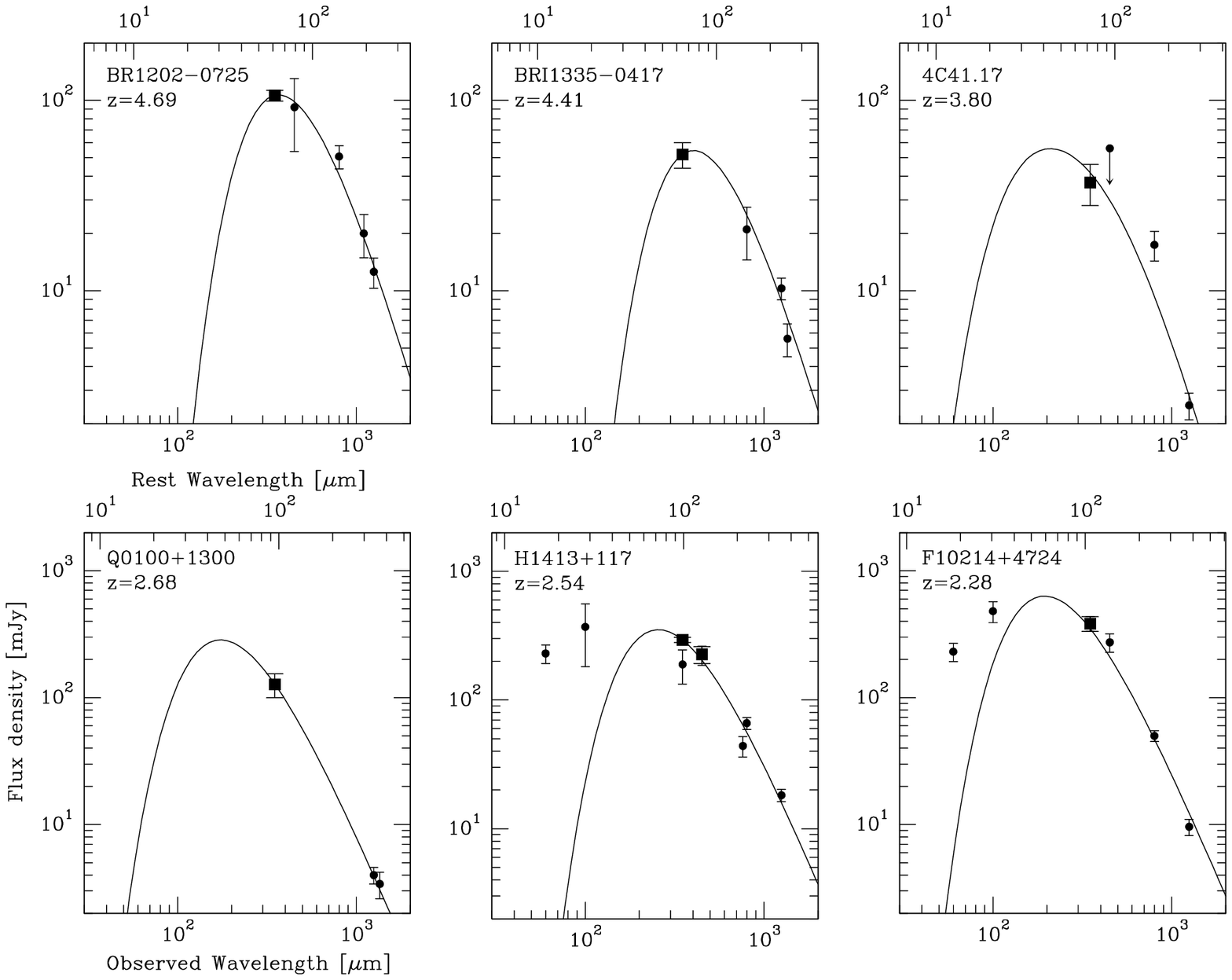, width=5in,angle=-90}}
\caption[]{Spectral energy distributions of six of the high-redshift objects
  discussed in this paper. The 350$\,\mu$m fluxes, shown as squares, are
  the measurements made with SHARC at the CSO. Fluxes taken from the
  literature are shown as circles - see Tables 1 for references.  
  The observed fluxes at
  $\lambda_{\rm obs}\geq 350\,\mu$m were approximated by grey body
  spectra, shown as solid lines, assuming an emissivity index $\beta =
  1.5$. The dust temperature results from a non-linear least-squares
  fit to the data - see text.
\label{seds}}
\end{figure*}

A greybody was fit to the data points $\rm S_\nu$, for wavelengths of
350$\,\mu$m and longward, as a function of the rest frequency
$\nu=\nu_{\rm obs}/(1+z)$, of the form
\begin{equation}
\rm S_\nu\,=\,B_\nu \Omega [1-\exp(-\tau)] 
{\rm ~with~ }\tau\,=\,(\nu/\nu_0)^\beta,
\end{equation}
where $\nu_0 = 2.4 \, {\rm THz}$ is the critical frequency at which
the source becomes optically thin and $\Omega$ is the solid angle of
emission. The shape of the fitted greybody is very weakly dependent on
the value of $\nu_0$ (\cite{Hughes}). The data were each weighted by
their statistical errors in the $\chi^2$ minimization.  This yields
the dust temperature, dust mass (following Hildebrand (1983), using a
dust mass emission coefficient at $\nu_0$ of $\rm 1.9~m^2~kg^{-1}$),
and luminosity of the sources.  When $\beta$ is considered as a free
parameter, we find that the average value of $\beta=1.5\pm0.2$ for the
detected sources.  The fits shown in Figure~\ref{seds} assume an
emissivity index of $\beta = 1.5$.  We estimated the $1\,\sigma$
uncertainty in the temperature by examining the $\chi_\nu^2$
hypersurface in the range $1\leq \beta\leq 2$, similarly to the method
of Hughes et al. (1993). To evaluate the uncertainties associated with
the mass and luminosity, derived from the fitted temperature and
$\beta=1.5$, we used the maximum and minimum values of the mass and
luminosity which are compatible with the data plus or minus the
statistical error. No lensing amplification was taken into account.
The temperature, dust mass and luminosities derived under these
assumptions are given in Table~\ref{detections}.  Two of the sources
with upper limits have 1.25mm detections (\cite{Omont96b}), which,
together with the 350$\,\mu$m data, yields an upper limit to their
temperature.  For Q$\,0842+3431$, we find that $\rm T_{dust}<40$K
while for BR$\,1117-1329$ a limit of $\rm T_{dust}<60K$ is found.  If
the dust is at the temperature limit, these quasars have dust masses
$< 10^8 M_\odot$.  For the other sources, an estimate of the maximum
luminosity has been given under the assumption that each object has a
temperature of 50$\,$K and an emissivity index of $\beta=1.5$.

The total luminosity is probably underestimated, since a large
luminosity contribution from higher temperature dust cannot be ruled
out for most sources.  However, in the case of H$\,$1413+117 and
IRAS~F10214+4724, the available IRAS data allow us to fit an
additional warm component.  For IRAS~F10214+4724, the cold component
model carries roughly 60\% of the total luminosity; in the case of
H$\,$1413+117, which has a hotter mid-IR spectrum, the total
luminosity is underestimated by a factor of 3.  Under the assumption
that the majority of the luminosity is carried by the cold component
(Table~\ref{detections}), the median luminosity-to-mass ratio is
around 100 $L_\odot/M_\odot$, assuming a gas-to-dust ratio of $\sim
500$ similar to that of IRAS~F10214+4724 and H$\,$1413+117
(\cite{Downes92}; \cite{Barvainis95}) or of ultraluminous infrared
galaxies (ULIRGs), i.e. $\rm 540\pm290$ (\cite{Sanders91}).  The peak
emission in the rest frame is found to be in the wavelength range $\rm
\lambda_{peak} \sim 60-80 \,\mu$m (Figure~\ref{seds}) implying dust
temperatures of 40-60$\,$K (Table~\ref{detections}). These
temperatures are nearly a factor of two lower than previously
estimated for ultraluminous sources (e.g. \cite{Chini94}). If the
temperature range we find is typical for the cold component of highly
redshifted objects, multi-band photometric studies in the
submillimeter/FIR, such as planned with FIRST, will provide reasonably
accurate redshift estimates for the sources detected in deep field
surveys.

The global star formation rate in each of the detected sources can be
estimated using the relation of Thronson \& Telesco (1986): ${\rm SFR}
\sim \Psi \, 10^{-10} L_{\rm FIR}/L_{\odot} h_{100}^{-2}\, M_{\odot}
\rm ~yr^{-1}$ with $\rm \Psi \sim\,$0.8-2.1.  For our mean luminosity
of $1.7\times10^{13}\,h_{100}^{-2} \,L_\odot$, this yields a SFR of
$\sim 2000 \, M_{\odot} h_{100}^{-2} \rm ~yr^{-1}$ (uncorrected for
lensing) if all the submillimeter flux is from a starburst component.
If we assume a final stellar mass of $\sim2\times10^{12} M_\odot$, a
value appropriate to a giant elliptical like M87 (\cite{Okazaki84}),
then the timescale for formation in a single massive starburst is
$\simeq 10^9h_{100}^2\,$yr.  Given the large mass of dust already
present in these quasars, a substantial amount of this star formation
must already have occurred.  For the most distant quasars, where the
age of the universe is similar to the derived formation timescale,
this implies a very high redshift ($z\gtrsim5$) for the era of initial
star formation, in agreement with models of high redshift Lyman-alpha
emitters (\cite{Haiman98}).

\begin{center} {\it Acknowledgments} \end{center}

The CSO is funded by the NSF under contract AST96-15025.  We thank
T.R. Hunter for help with the fitting/derivation programming and D.
Downes for helpful comments.  One of us (P.C.) acknowledges
financial supports from INSU (Programmes Grands T\'elescopes
Etrangers) and PCMI.


\begin{thebibliography}{}
  
\bibitem[Alloin et al. 1997]{Alloin97} Alloin, D., Guilloteau, S.,
        Barvainis, R., \& Tacconi, L. 1997, A\&A, 321, 24
\bibitem[Barger et al. 1998]{Barger98} Barger, A.J., Cowie,
        L.L., Sanders, D.B., Fulton, E., Taniguchi, Y., Sato, Y.,
        Kawara, K., Okuda, H.  1998, Nature, 394, 248
\bibitem[Barvainis, Antonucci, \& Coleman 1992]{Barvainis92} Barvainis, R.,
        Antonucci, R., \& Coleman, P. 1992, ApJ, 399, L19
\bibitem[Barvainis et al. 1995]{Barvainis95} Barvainis, R.,
        Antonucci, R., Hurt, T., Coleman, P., \& Reuter, H.-P.  1995, ApJ,
        451, L9
\bibitem[Barvainis et al. 1998]{barvainis4} Barvainis, R., Alloin,
        D., Guilloteau, S., \& Antonucci, R.  1998, ApJ, 492, L13
\bibitem[Benford et al. 1998]{Benford98} Benford, D.J., Cox, P.,
        Omont, A., \& Phillips, T.G. 1998, BAAS, 192, 11.04
\bibitem[Chambers, Miley \& van Breugel 1990]{CMB90} Chambers,
        K.C., Miley, G.K., \& van Breugel, W.J.M. 1990, ApJ, 363, 21
\bibitem[Chini \& Kr\"ugel 1994]{Chini94}
        Chini, R. \& Kr\"ugel, E., 1994, A\&A, 288, L33
\bibitem[Downes et al. 1992]{Downes92} Downes, D., Radford, S.J.E.,
        Greve, A., Thum, C., Solomon, P.M., \& Wink, J.E. 1992, ApJ, 398, L25
\bibitem[Downes, Solomon, \& Radford 1995]{Downes95}Downes, D.,
        Solomon, P.M., \& Radford, S.J.E. 1995, ApJ, 453, L65
\bibitem[Downes \& Solomon 1998]{Downes98} Downes, D. \& Solomon,
        P.M. 1998, ApJ, 507, 615
\bibitem[Fixsen et al. 1998]{Fixsen98}Fixsen, D.J., Dwek, E.,
        Mather, J.C., Bennett, C.L. \& Shafer, R.A., 1998, ApJ, 508, 123
\bibitem[Foltz et al. 1987]{Foltz87} Foltz, C.B., Chaffee, F.H.,
        Hewett, P.C., McAlpine, G.M., Turnsheck, D.A., Weymann, R.J.,
        \& Anderson, S.F. 1987, AJ, 94, 1423
\bibitem[Guilloteau et al. 1997]{Guilloteau97} Guilloteau, S., Omont,
        A., McMahon, R.G, Cox, P., \& Petitjean, P. 1997, A\&A, 328, L1
\bibitem[Guilloteau et al. 1999]{Guilloteau99} ---. 1999, in preparation
\bibitem[Haas et al. 1998]{Haas98}
        Haas, M., Chini, R., Meisenheimer, K., Stickel, M., Lemke, D., Klaas,
        U. \& Kreysa, E. 1998, ApJ, 503, L109
\bibitem[Haiman \& Spaans 1999]{Haiman98} Haiman, Z. \& Spaans, M.
        1999, ApJ, in press; {\tt astro-ph/9809223}
\bibitem[Hauser et al. 1998]{Hauser98} Hauser, M.G., et al., 1998,
        ApJ, 508, 25
\bibitem[Hazard et al. 1984]{Haz94} Hazard, C., Morton, D.C.,
        Terlevich, R., \& McMahon, R.G. 1984, ApJ, 282, 33
\bibitem[Hewett et al. 1995]{Hewett95} Hewett, P.C., Foltz, C.B.,
        \& Chaffee, F.H. 1995, AJ, 109, 1498
\bibitem[Hildebrand 1983]{Hildebrand83} Hildebrand, R.H. 1983, QJRAS,
        24, 267
\bibitem[Hughes et al. 1993]{Hughes} Hughes, D.H., Robson, E.I.,
        Dunlop, J.S. \& Gear, W.K. 1993, MNRAS, 263, 607
\bibitem[Hughes, Dunlop, \& Rawlings 1997]{Hughes97} Hughes, D.H.,
        Dunlop, J.S., \&  Rawlings, S. 1997, MNRAS, 289, 766
\bibitem[Hughes et al. 1998]{Hughes98} Hughes, D. H., et al., 1998,
        Nature, 394, 241
\bibitem [Hunter, Benford, \& Serabyn 1996]{Hunter96} Hunter, T.R.,
        Benford, D.J. \& Serabyn, E.  1996, PASP, 108, 1042
\bibitem[Isaak et al. 1994]{Isaak} Isaak, K.G., McMahon, R.G., Hills,
        R.E., \& Withington, S. 1994, MNRAS, 269, L28
\bibitem[Ivison (1995)]{Ivison95} Ivison, R.J. 1995, MNRAS, 275, L33
\bibitem[Ivison et al. 1998]{Ivison98} Ivison, R.J., Smail, I., Le Borgne,
        J.-F., Blain, A.W., Kneib, J.-P., B\'ezecourt, J., Kerr, T.H., \&
        Davies, J.K. 1998, MNRAS, 298, 583
\bibitem[Kawara et al. 1998]{Kawara98} Kawara, K., et al., 1998, A\&A, 336, L9
\bibitem[Kennefick et al. 1995]{Kennefick95} Kennefick, J. D., De
        Carvalho, R. R., Djorgovski, S. G., Wilber, M. M., Dickson, E. S.,
        Weir, N., Fayyad, U. \& Roden, J. 1995, AJ, 110, 78
\bibitem[Lewis et al. 1998]{Lewis98}
        Lewis, G.F., Chapman, S.C., Ibata, R.A., Irwin, M.J. \& Totten, E.J.
        1998, ApJ, 505, L1
\bibitem[Lilly et al. 1999]{Lilly99}
        Lilly, S.J., Eales, S.A., Gear, W.K.P., Hammer, F., Le F\`evre, O.,
        Crampton, D., Bond, J.R. \& Dunne, L. 1999, ApJ, in press;
        {\tt astro-ph/9901047}
\bibitem[McMahon et al. 1994]{McMahon94} McMahon, R.G, Omont, A.,
        Bergeron, J., Kreysa, E., \& Haslam, C.G.T. 1994, MNRAS, 267, L9
\bibitem[Okazaki \& Inagaki 1984]{Okazaki84} Okazaki, A. T. \&
        Inagaki, S. 1984, PASJ, 36, 17
\bibitem[Omont et al. 1996a]{Omont96a} Omont, A., Petitjean, P.,
        Guilloteau, S., McMahon, R.G, Solomon, P.M., \& P\'econtal, E.
        1996a, Nature, 382, 428
\bibitem[Omont et al. 1996b]{Omont96b} Omont, A., McMahon, R.G, Cox,
        P., Kreysa, E., Bergeron, J., Pajot, F., \& Storrie-Lombardi, L.J.
        1996b, A\&A, 315, 1
\bibitem[Phillips 1997]{Phillips}Phillips, T.G.  1997, in {\it The Far
        Infrared and Submillimetre Universe}, ESA SP-401, p. 223, A.
        Wilson, ed.
\bibitem[Puget et al. 1996]{Puget96} Puget, J.-L., Abergel, A.,
        Bernard, J.-P., Boulanger, F., Burton, W.B., Desert, F.-X. \&
        Hartmann, D., 1996, A\&A, 308, L5
\bibitem[Puget et al. 1999]{Puget98} Puget, J.-L. et al., 1999, A\&A,
        in press
\bibitem[Rowan-Robinson et al. 1991]{RR91} Rowan-Robinson, M., et al.,
        1991, Nature, 351, 719
\bibitem[Rowan-Robinson et al.  1993]{RR93} Rowan-Robinson, M., et
        al., 1993, MNRAS, 261, 513
\bibitem[Sanders, Scoville, \& Soifer 1991]{Sanders91} Sanders, D.B.,
        Scoville, N.Z., \& Soifer, B.T. 1991, ApJ, 370, 158
\bibitem[Sargent et al. 1989]{Sargent89} Sargent, W.L.W., Steidel, C.,
        \& Boksenberg, A. 1989, ApJSS, 69, 703
\bibitem[Schneider, Schmidt \& Gunn (1989)]{Schneider89}
        Schneider, D.P., Schmidt, M., \& Gunn, J. 1989, AJ, 98, 1507
\bibitem[Schneider, Schmidt \& Gunn (1994)]{Schneider94}
        Schneider, D.P., Schmidt, M., \& Gunn, J. 1994, AJ, 107, 1245
\bibitem[Steidel \& Sargent 1991]{Steidel91} Steidel, C.C. \& Sargent, W.L.W.
        1991, ApJ, 382, 433
\bibitem[Storrie-Lombardi et al. (1996)]{SL96} Storrie-Lombardi,
        L.J., McMahon, R.G., Irwin, M.J., \& Hazard, C.  1996, ApJ,
        468, 121
\bibitem[Thronson \& Telesco 1986]{Thronson86} Thronson, H., \&
        Telesco, C. 1986, ApJ, 311, 98 
\bibitem[Wang et al. 1996]{Wang96}  Wang, N., et al., 1996, Applied
        Optics 35, 6629

\end{thebibliography}
\end{document}